\documentclass[12pt]{article}
\usepackage[centertags]{amsmath}
\usepackage{amsfonts}
\usepackage{amssymb}
\usepackage{amsthm}
\usepackage{newlfont}
\usepackage{hyperref}

\numberwithin{equation}{section}

\newtheorem{theorem}{Theorem}

\newtheorem{definition}{Definition}

\begin{document}

\title{The G-dynamics of the QCPB theory}

\author{ Gen  WANG \footnote{School of Mathematical Sciences, Xiamen University,
     Xiamen, 361005, P.R.China. Email: wanggen@zjnu.edu.cn}}

\date{}

\maketitle

\begin{abstract}
In this paper, we further study the G-dynamics newly emerged in the covariant dynamics defined by the  quantum covariant Poisson bracket (QCPB) theory. We propose three new operators based on the G-dynamics, we find that the non-Hermitian operators are inevitable to appear in the framework of the QCPB. Then we seek the precise formulations of the G-dynamics to calculate the practical quantum problems, the G-dynamics of three types in terms of different Hamiltonian are discovered.
 The geometrinetic energy operator for quantum harmonic oscillator as an application is further calculated.

\end{abstract}

\tableofcontents

\section{Introduction}

\subsection{The Schr\"{o}dinger equation}
 In quantum mechanics, the form of the Schr\"{o}dinger equation depends on the physical situation. The most general form is the time-dependent Schr\"{o}dinger equation, which gives a description of a system evolving with time \cite{1,2,4}
 \begin{equation}\label{eq2}
   \sqrt{-1}\hbar \frac{\partial \psi }{\partial t}={{\hat{H}}^{\left( cl \right)}}\psi={-\frac {\hbar ^{2}}{2m }}\nabla ^{2}\psi+V\psi
 \end{equation}
where $\partial/\partial t$ symbolizes a partial derivative with respect to time $t$, $\psi$  is the wave function of the quantum system,
\begin{equation}\label{eq15}
  {{\hat{H}}^{\left( cl \right)}}=-\frac{{{\hbar }^{2}}}{2m}{{\nabla }^{2}}+V
\end{equation}
is the Hamiltonian operator in terms of the coordinates.   The time-dependent Schr\"{o}dinger equation described above predicts that wave functions can form standing waves, called stationary states.  Stationary states can also be described by a simpler form of the Schr\"{o}dinger equation, the time-independent Schr\"{o}dinger equation
$${\displaystyle{{{\hat{H}}^{\left( cl \right)}}} \psi =\left({-\frac {\hbar ^{2}}{2m }}\nabla ^{2}+V\right)\psi ={{E}^{\left( cl \right)}}\psi }$$
The energy and momentum operators, which are respectively shown as
\begin{equation}\label{eq4}
  {{\hat{H}}^{\left( cl \right)}}=\sqrt{-1}\hbar\frac{\partial}{\partial t}\,,\quad \hat{p}^{\left( cl \right)} = -\sqrt{-1}\hbar\nabla
\end{equation}
The Schr\"{o}dinger equation in its general form is shown as \eqref{eq2}.

With the non-universal validity of the quantum Poisson bracket that has been replaced by the complete quantum covariant Poisson bracket theory. In this paper, on the foundation of the Schr\"{o}dinger equation, we mainly explore the properties of the G-dynamics in the covariant dynamics defined by the QCHS in QCPB, we obtain the more specific expressions of the G-dynamics in terms of the coordinates to make the G-dynamics practical in the quantum mechanics as it develops.

\section{Generalized geometric commutator and geometric anticommutator}

Let $a$ and $b$ be any two operators, their commutator is formally defined by
${\displaystyle [a,b]_{cr}=ab-ba}$.
Note that the operator for commutator can be any mathematical form to be appeared for the calculation, such as a function, vector, differential operator, partial differential operator, even a number in a number field, and so on, it can be arbitrarily chosen according to our needs.

Let $a$ and $b$ be any elements of any algebra, or any operators, their generalized geometric commutator is formally defined by the following.
\begin{definition}[GGC] \cite{3}
A generalized geometric commutator (GGC) of arity two $a,b$ is formally given by
$$\left[ a,b \right]={{\left[ a,b \right]}_{cr}}+G\left( s,a,b \right)$$
The geomutator is $$G\left( s,a,b \right)=a{{\left[ s,b \right]}_{cr}}-b{{\left[ s,a \right]}_{cr}}$$ satisfying  $G\left( s,a,b \right)=-G\left( s,b,a\right)$, where $s$ is a geometric structure function given by domain.
\end{definition}

Some properties of the geomutator are given by
\begin{align}
  & G\left( s,a+b,c+d \right)=G\left( s,a,c \right)+G\left( s,b,d \right)+G\left( s,b,c \right)+G\left( s,a,d \right) \notag\\
 & G\left( s,a+b,c \right)=G\left( s,a,c \right)+G\left( s,b,c \right)\notag \\
 & G\left( s,a,c+d \right)=G\left( s,a,c \right)+G\left( s,a,d \right)\notag
\end{align}
for operators $a,b,c,d$.

Obviously, in some sense, the anticommutator also needs to be generalized as the commutator does,  Analogously, denote by ${{\left\{ a,b \right\}}_{ir}}=ab+ba$ the anticommutator.
\begin{definition}
 The geometric anticommutator (GAC) of any two elements $a$ and $b$ is defined by
  \[\left\{ a,b \right\}={{\left\{ a,b \right\}}_{ir}}+Z\left(s, a,b \right)\]
  where $Z\left(s,a,b \right)=Z\left(s, b,a\right)$ is called anti-geomutator, $s$ is geometric function.
\end{definition}
As the definition above stated,

\begin{definition}
The anti-geomutator can be taken as
\[Z\left(s,a,b \right)=\left( a:s:b \right)+{{\left\{ a,b \right\}}_{ir}}s\]
where $\left( a:s:b \right)=asb+bsa$, and $s$ is the geometric function created by the environment.
\end{definition}
As a result of the symmetry of anti-geomutator, geometric anticommutator then follows the symmetry $\left\{ a,b \right\}=\left\{ b,a \right\}$.
 Actually, the anti-geomutator can be expressed in the form
\begin{align}
  Z\left( s,a,b \right)& =a{{\left\{ s,b \right\}}_{ir}}+b{{\left\{ s,a \right\}}_{ir}}=\left( a:s:b \right)+{{\left\{ a,b \right\}}_{ir}}s \notag
\end{align}
As seen, this form is very similar to the quantum geometric bracket, this is why we need to generalize the anticommutator. There has a clear property of the anti-geomutator given by
\begin{align}
 Z\left( s,a+b,c+d \right) &=\left( a+b \right){{\left\{ s,c+d \right\}}_{ir}}+\left( c+d \right){{\left\{ s,a+b \right\}}_{ir}}\notag \\
 & =Z\left( s,a,c \right)+Z\left( s,a,d \right)+Z\left( s,b,c \right)+Z\left( s,b,d \right) \notag
\end{align}
for elements $a,b,c,d$.

\subsection{Quantum covariant Poisson bracket and quantum geobracket (QGB)}
As \cite{3} stated, the quantum covariant Poisson bracket (QCPB) is defined by generalized geometric commutator (GGC) while quantum geometric bracket (QGB) is given based on the geomutator.  More precisely,

\begin{definition}[QCPB] \cite{3} \label{d1}
  The QCPB is generally defined as
 \[\left[ \hat{f},\hat{g} \right]={{\left[ \hat{f},\hat{g} \right]}_{QPB}}+G\left( s,\hat{f},\hat{g} \right)\]in terms of quantum operator $\hat{f},~\hat{g}$, where $$G\left(s, \hat{f},\hat{g} \right)=\hat{f}{{\left[ s,\hat{g} \right]}_{QPB}}-\hat{g}{{\left[ s,\hat{f} \right]}_{QPB}}=-G\left(s, \hat{g},\hat{f} \right)$$ is called quantum geometric bracket, where geometric
potential function  $s$ represents the globally condition of space.
\end{definition}It is zero if and only if $\hat{f}$ and $\hat{g}$ covariant commute, i,e. $\left[ \hat{f},\hat{g} \right]=0$.  It is remarkable to see that the QCPB representation admits a dynamical geometric bracket formula on the manifold.

Let's assert the role of the geometric scalar potential function s only represents the attributes of the manifold space itself, it means that $s$ is a geometric structure function given by domain based on the generalized geometric commutator in definition \ref{d1}, it means that geometric potential function $s$  is only determined by the environment, or spacetime, or manifolds, the domain, ect, from this viewpoint, the environment joins the physical process, the influence of the environment now based on new theory can't be ignored, it's naturally necessary to be considered in a physical process.
\begin{definition} \cite{3}
 The covariant equilibrium equation is given by $\left[ \hat{f},\hat{g} \right]=0$, i.e,
  \[{{\left[ \hat{f},\hat{g} \right]}_{QPB}}+G\left( s, \hat{f},\hat{g} \right)=0\]for operators $\hat{f},~\hat{g}$.
\end{definition}

\subsection{Quantum covariant Hamiltonian system}
This section will give the covariant dynamics which contains two different sub-dynamics: the generalized Heisenberg equation and G-dynamics  \cite{3}. It tells that the generalized Heisenberg equation has improved the classical Heisenberg equation by considering the quantum geometric bracket. Firstly, let's give the definition of quantum covariant Hamiltonian system (QCHS) based on the QCPB as definition \ref{d1} defined.

In the beginning, let us start with the QCPB by taking $\hat{g}=\hat{H}$ into consideration, then it formally yields below definition.
\begin{definition}[QCHS] \cite{3}\label{d10}
   The QCHS defined by QCPB in terms of quantum operator $\hat{f},~\hat{H}$ is generally given by
 \[\left[ \hat{f},\hat{H} \right]={{\left[ \hat{f},\hat{H} \right]}_{QPB}}+G\left(s, \hat{f},\hat{H} \right)\]where $$G\left(s, \hat{f},\hat{H} \right)=\hat{f}{{\left[ s,\hat{H} \right]}_{QPB}}-\hat{H}{{\left[ s,\hat{f} \right]}_{QPB}}$$ is quantum geobracket .
\end{definition}
Manifestly, this operator $\hat{f}$ covariantly commutes with $\hat{H}$.  By our construction, the QCHS defined by the QCPB is a transition for the further development of a complete and covariant theory that can naturally generalize the Heisenberg equation. Since $\hat{f}$ is an observable, $s$ is assumed to be real function as well. In this way, we have successfully given a complete description of the Heisenberg equation.
\begin{definition}\label{d4} \cite{3}
 The covariant equilibrium equation is given by $\left[ \hat{f},\hat{H} \right]=0$, i.e,
  \[{{\left[ \hat{f},\hat{H} \right]}_{QPB}}+G\left( s, \hat{f},\hat{H} \right)=0\]for operators $\hat{f},~\hat{H}$, then $\hat{f}$ is called quantum covariant conserved quantity.
\end{definition}

Furthermore, there is a certain case given by
$\left[ \hat{H},\hat{H} \right]=0$,
when $\hat{f}=\hat{H}$ based on definition \ref{d4}. It is clear to see that $\hat{H}$ is a quantum covariant conserved quantity.

\subsection{Covariant dynamics, generalized Heisenberg equation, G-dynamics}
In this section, we will briefly review the entire theoretical framework of quantum covariant Hamiltonian system defined by the quantum covariant Poisson bracket totally based on the paper \cite{3}.  More precisely,
the time covariant evolution of any observable $\hat{f}$ in the covariant dynamics is given by both the generalized Heisenberg equation of motion and G-dynamics.

The covariant dynamics is given by $\frac{\mathcal{D}\hat{f}}{dt}=\frac{1}{\sqrt{-1}\hbar }\left[ \hat{f},\hat{H} \right]$, it includes the generalized Heisenberg equation, G-dynamics can be formally formulated as
\begin{description}
  \item[The generalized Heisenberg equation:] $$\frac{d\hat{f}}{dt}=\frac{1}{\sqrt{-1}\hbar }{{\left[ \hat{f},\hat{H} \right]}_{QPB}}-\frac{1}{\sqrt{-1}\hbar }\hat{H}{{\left[ s,\hat{f} \right]}_{QPB}}$$
  \item[G-dynamics:] $\hat{w}=\frac{1}{\sqrt{-1}\hbar }{{\left[ s,\hat{H} \right]}_{QPB}}$.
\end{description}respectively, where $\frac{\mathcal{D}}{dt}=\frac{d}{dt}+\hat{w}$ is covariant time operator, and
${{\hat{H}}}$ is the Hamiltonian and $[\cdot,\cdot]$ denotes the GGC of two operators.

Note that the G-dynamics in terms of $\hat{H}^{\left( cl \right)}$ describes the quantum rotation of the manifolds space in a certainty quantum system, its eigenvalues represent the frequency spectrum.
As the formulation of G-dynamics shows, it's obvious to see that the imaginary geomenergy can be accordingly defined based on the G-dynamics,  by using this new concept, we can better give a presentation for the covariant dynamics, ect.
\begin{definition} \label{d2} \cite{3}
Based on the G-dynamics $\hat{w}$, then the imaginary geomenergy is induced by ${{E}^{\left( \operatorname{Im} \right)}}\left( \hat{w} \right)=\sqrt{-1}\hbar \hat{w}={{\left[ s,\hat{H} \right]}_{QPB}}$.
\end{definition} It is clear that the covariant dynamics leads us to a complete quantum system in which the state of the system is represented by two separated dynamic quantum system, the covariant dynamics can be totally pictured in the form below
\begin{align}
 \sqrt{-1}\hbar \frac{\mathcal{D}}{dt}\hat{f} &=\left[ \hat{f},\hat{H} \right] ={{\left[ \hat{f},\hat{H} \right]}_{QPB}}+G\left( s, \hat{f},\hat{H} \right) \notag\\
 & =\sqrt{-1}\hbar \frac{d}{dt}\hat{f}+\sqrt{-1}\hbar \hat{f}\hat{w} \notag
\end{align}
With the imaginary geomenergy defined above, then covariant dynamics is rewritten in the form
\[\sqrt{-1}\hbar \frac{\mathcal{D}}{dt}\hat{f}=\left[ \hat{f},\hat{H} \right]=\sqrt{-1}\hbar \frac{d}{dt}\hat{f}+\hat{f}{{E}^{\left( \operatorname{Im} \right)}}\left( \hat{w} \right)\]
As a consequence of the imaginary geomenergy, we can say that imaginary geomenergy is a new kind of Hamiltonian operator.

\subsection{The QCPB for quantum harmonic oscillator}
In this section, we simply review some basic results given by the paper
\cite{3} in QCPB.  As quantum mechanics illustrated, the commutator relations may look different than in the Schr\"{o}dinger picture, because of the time dependence of operators. The time evolution of those operators depends on the Hamiltonian of the system. Considering the one-dimensional quantum harmonic oscillator,
\begin{equation}\label{eq3}
  \hat{H}^{\left( cl \right)}={\frac  {{{\hat{p}}}^{{\left( cl \right)2}}}{2m}}+{\frac  {m\omega ^{{2}}x^{{2}}}{2}}
\end{equation}
The evolutions of the position and momentum operators are given by:
\begin{equation}\label{eq5}
  {d \over dt}x(t)={\sqrt{-1} \over \hbar }[\hat{H}^{\left( cl \right)},x(t)]_{QPB}={\frac  {{{\hat{p}}}^{{\left( cl \right)}}}{m}}
\end{equation}\[{d \over dt}{{\hat{p}}}^{{\left( cl \right)}}(t)={\sqrt{-1}\over \hbar }[\hat{H}^{\left( cl \right)},{{\hat{p}}}^{{\left( cl \right)}}(t)]_{QPB}=-m\omega ^{{2}}x\]
As for the application of the QPB or the commutation, there are many fields including the physics and mathematics, and so on.

As a certainly example,  we will now start by briefly reviewing the quantum mechanics of a one-dimensional quantum harmonic oscillator \eqref{eq3}, and see how the QCPB can be incorporated using GGC in the covariant quantization procedure.
With the Hamiltonian given by \eqref{eq3}, let's use the QCPB to recalculate the \eqref{eq5}, we can see how differences emerge. More specifically, the covariant dynamics in terms of the position reads
\[\frac{\mathcal{D}}{dt}x\left( t \right)=\frac{\sqrt{-1}}{\hbar }\left[ \hat{H}^{\left( cl \right)},x\left( t \right) \right]=\frac{{{\hat{p}}^{\left( cl \right)}}\left( t \right)}{m}+x\left( t \right){{\hat{w}}^{\left( cl \right)}}\]
By direct computation, the G-dynamics in terms of $\hat{H}^{\left( cl \right)}$ is given by
\begin{align}\label{eq6}
 {{\hat{w}}^{\left( cl \right)}}&=-\sqrt{-1}{{\left[ s,{{\hat{H}}^{\left( cl\right)}}\right]}_{QPB}}/\hbar=\frac{\sqrt{-1}}{\hbar }\frac{{{\hat{p}}^{\left( cl \right)2}}s+2{{\hat{p}}^{\left( cl \right)}}s{{\hat{p}}^{\left( cl \right)}}}{2m}\\
 &=b_{c}\left( 2u\frac{d}{dx}+u_{x} \right)\notag
\end{align}where $b_{c}=-\frac{\sqrt{-1}\hbar }{2m}$, and $u=\frac{ds}{dx}$, $u_{x}=\frac{{{d}^{2}}}{d{{x}^{2}}}s$ are used.
Furthermore, it gets $\sqrt{-1}\hbar {{{\hat{w}}}^{\left( cl \right)}}={{\left[ s,{{\hat{H}}^{\left( cl\right)}}\right]}_{QPB}}$, then it leads to
$$\sqrt{-1}\hbar {{\hat{w}}^{\left( cl \right)}}\psi=\frac{{{\hbar }^{2}}}{2m}\left( 2u{{\psi }_{x}}+\psi {{u}_{x}} \right)$$
And the generalized Heisenberg equation with respect to $x$ follows
\begin{align}
 \frac{d}{dt}x\left( t \right) & =\frac{\sqrt{-1}}{\hbar }{{\left[ \hat{H}^{\left( cl \right)},x\left( t \right) \right]}_{QPB}}+\frac{\sqrt{-1}}{\hbar }\hat{H}^{\left( cl \right)}{{\left[ s,x\left( t \right) \right]}_{QPB}} \notag\\
 & =\frac{{{\hat{p}}^{\left( cl \right)}}\left( t \right)}{m}\notag
\end{align}where ${{\left[ s,x\left( t \right) \right]}_{QPB}}=0$.
In the same way,  the covariant dynamics for the classical momentum operator is
\begin{align}
  \frac{\mathcal{D}}{dt}{{{\hat{p}}}^{\left( cl \right)}}\left( t \right)&=\frac{\sqrt{-1}}{\hbar }\left[\hat{H}^{\left( cl \right)},{{{\hat{p}}}^{\left( cl \right)}}\left( t \right) \right] \notag\\
 & =-m{{\omega }^{2}}x-\hat{H}^{\left( cl \right)}u+{{\hat{p}}^{\left( cl \right)}}\left( t \right){{\hat{w}}^{\left( cl \right)}} \notag
\end{align}
Accordingly, the generalized Heisenberg equation in terms of the ${{\hat{p}}}^{{\left( cl \right)}}$ appears
\begin{align}
 \frac{d}{dt}{{{\hat{p}}}^{\left( cl \right)}}\left( t \right) &=\frac{\sqrt{-1}}{\hbar }{{\left[ \hat{H}^{\left( cl \right)},{{{\hat{p}}}^{\left( cl \right)}}\left( t \right) \right]}_{QPB}}+\frac{\sqrt{-1}}{\hbar }\hat{H}^{\left( cl \right)}{{\left[ s,{{{\hat{p}}}^{\left( cl \right)}}\left( t \right) \right]}_{QPB}} \notag\\
 & =-m{{\omega }^{2}}x-\hat{H}^{\left( cl \right)}u \notag
\end{align}
where $\sqrt{-1}\hbar u={{\left[ s,{{{\hat{p}}}^{\left( cl \right)}}\left( t \right) \right]}_{QPB}}$.

As for the quantum operator \eqref{eq6}, we can do some transformations in terms of the function $u$, more precisely, some examples are given as follows:

If let $u=mx$ be given, then when it brings to \eqref{eq6} for a replacement,  then it gets the Berry-Keating’s Hamiltonian expressed as
${{\hat{H}}^{\left( \text{bk} \right)}}=-\sqrt{-1}\hbar \left( x\frac{d}{dx}+\frac{1}{2} \right)$.

If let  $u=m\omega_{0}x/\hbar$, where $\omega_{0}$ is a constant frequency, then ${{\hat{H}}^{\left( \text{bk} \right)}}/\hbar=-\sqrt{-1}\omega_{0}\left( x\frac{d}{dx}+\frac{1}{2} \right)$.

If let $u=x$ be given, then $u_{x}=1$,  then it gets
${{\hat{L}}}=2b_{c} \left( x\frac{d}{dx}+\frac{1}{2} \right)$.

Similarly, if we set $u={{a}_{0}}{{x}^{-1}}$, where ${{a}_{0}}$ is a constant, then
the \eqref{a1} in this case becomes
$\hat{L}=2{{a}_{0}}{{b}_{c}}{{x}^{-2}}\left( x\frac{d}{dx}-\frac{1}{2} \right)$.

If let $u={{a}_{2}}{{x}^{-2}}$, then $\hat{L}=2{{a}_{2}}{{b}_{c}}{{x}^{-3}}\left( x\frac{d}{dx}-1 \right)$.

If we choose $u={{a}_{4}}{{x}^{-2}}+{{a}_{1}}x$, then it obtains
$$\hat{L}=2{{a}_{4}}{{b}_{c}}{{x}^{-3}}\left( x\frac{d}{dx}-1 \right)+2{{a}_{1}}{{b}_{c}}\left( x\frac{d}{dx}+\frac{1}{2} \right)$$
 where ${{a}_{1}},{{a}_{4}}$ are the constants.

If we set $u={{a}_{3}}{{x}^{2}}$, then it obtains
$\hat{L}={{a}_{3}}{{b}_{c}}{{x}}\left( x\frac{d}{dx}+2 \right)$.

In particular, if we let $u_{x}=0$, then $\hat{L}=2{{a}_{5}}{{b}_{c}}\frac{d}{dx}={{a}_{5}}\frac{{{\hat{p}}^{\left( cl \right)}}}{m}$, where $\hat{p}^{(cl)}=-\sqrt{-1}\hbar\frac{d}{dx}$ is classical momentum operator in one dimension, furthermore, ${a}_{5}=m$ leads to $\hat{L}=\hat{p}^{(cl)}$.
As the different transformations shows above, the all results almost take the similar mode, this implies that the peculia feature of the quantum operator \eqref{eq6} holds intricately.

\subsection{Geomentum operator}
\begin{definition}\label{d5} \cite{3}
  Let $M$ be a smooth manifold represented by geometric
potential function $s$, then geomentum operator is defined as
  $$\hat{p}=-\sqrt{-1}\hbar D$$ where $D=\nabla +\nabla s$. The component is ${{\hat{p}}_{j}}=-\sqrt{-1}\hbar {{D}_{j}}$ in which ${{D}_{j}}={{\partial }_{j}}+{{\partial }_{j}}s$ holds, ${{\partial }_{j}}=\frac{\partial }{\partial {{x}_{j}}}$.
\end{definition}
Note that the geomentum operator is a revision of the classical momentum operator given by \eqref{eq4}.

\begin{theorem}[Geometric canonical quantization rules] \cite{3}
  Geometric equal-time canonical commutation relation is
  \[\left[ {{{x}_{i}}},\hat{{{p}_{j}}} \right]=\sqrt{-1}\hbar {{D}_{j}}{{x}_{i}}\]where  $\left[ \cdot ,~\cdot  \right]={{\left[ \cdot ,~\cdot  \right]}_{QPB}}+G\left(s,\cdot ,~\cdot  \right)$ is QCPB.
\end{theorem}
Geometric canonical commutation relation can be expressed in a specific form \[\left[ {{{x}_{i}}},\hat{{{p}_{j}}} \right]=\sqrt{-1}\hbar \left( {{\delta }_{ij}}+{{x}_{i}}u_{j}\right) \]where $u_{j}=\frac{\partial }{\partial {{x}_{j}}}s={{\partial }_{j}}s$.
In other words, it also can be rewritten as
$\left[ {{x}_{i}},\hat{{{p}_{j}}} \right]=\sqrt{-1}\hbar {{\theta }_{ij}}$, where
${{\theta}_{ij}}={{\delta }_{ij}}+{{x}_{i}}u_{j}$, and ${{\partial }_{j}}=\frac{\partial }{\partial {{x}_{j}}}$.

\section{G-dynamics}
In this section, G-dynamics as a new dynamical form appears in the quantum mechanics, we mainly focus on the G-dynamics as a part of covariant dynamics, that is, $\hat{w}=\frac{1}{\sqrt{-1}\hbar }{{\left[ s,\hat{H} \right]}_{QPB}}$.  As \cite{3} already stated, the G-dynamics is only induced by the geometric
potential function $s$, and meanwhile, it's independent to the other observables.  There is a clear fact of the G-dynamics taken as a different form by choosing different Hamiltonian operators, in one word, various Hamiltonian operators correspond to the multiple G-dynamics.
Therefore, for a given wave function $\psi$, the eigenvalue equation of operator is accordingly given by $\hat{w}\psi ={{w}}\psi$, where ${{w}}$ is the eigenvalue.
By using definition \ref{d2}, it gets corresponding imaginary geomenergy given by definition \ref{d2}, that is,
$${{E}^{\left( \operatorname{Im} \right)}}=\sqrt{-1}\hbar \hat{w}={{\left[ s,\hat{H} \right]}_{QPB}}$$ Thusly, it then has energy spectrum given by
${{E}^{\left( \operatorname{Im} \right)}}\psi=\sqrt{-1}{{E}^{\left( g \right)}}\psi $,  where ${E}^{\left( g \right)}=\hbar w$.
Obviously, we can see that the G-dynamics is completely determined by the geometric
potential function $s$, it implies that the G-dynamics represents the properties of the spatial manifolds, in other words, the spatial manifolds has abundant activities. As a result of this point, we need to seek more clues to unlock this quantum characters. The classical Hamiltonian operator \eqref{eq4} or \eqref{eq15} is the first to bear the brunt to be considered in above formula.

\subsection{D-operator, T-operator and G-operator}
In this section, the discussions on the most general form will be done,  we use the G-dynamics $\hat{w}$ to define the T-operator, and then the imaginary geomenergy is used with the classical Hamiltonian  operator \eqref{eq4} to construct the G-operator as a non-Hermitian operator
to deeply study some properties of the G-dynamics.

To start with the definition of the D-operator and the T-operator,
\begin{definition}\label{d6}
The D-operator is given by $\hat{{{D}_{t}}}={{\partial }_{t}}-\hat{w}$, and then T-operator is defined as
 $\hat{{{N}_{t}}}=\sqrt{-1}\hat{{{D}_{t}}}$,
  where $\hat{w}$ is the G-dynamics.
\end{definition}
Note that D-operator $\hat{{{D}_{t}}}={{\partial }_{t}}-\hat{w}$ is one of cases,  it means that there are two kinds of D-operators, that is another case $\hat{{{D}_{t}}}={{\partial }_{t}}+\hat{w}$.  More precisely, we take one of them into consideration as reality needed, we mainly focus on the plus case.
Its eigenvalue equation of D-operator is given by $\hat{{{D}_{t}}}\psi ={{D}_{t}}\psi$.

As a result of the definition \ref{d6}, that naturally leads to bond the imaginary geomenergy and the classical Hamiltonian operator \eqref{eq4} together to construct a new energy operator, accordingly, we can derive following G-operator.
\begin{definition}\label{d3}
The G-operator can be defined as
${{\hat{H}}^{\left( gr \right)}}=\hbar \hat{{{N}_{t}}}$.
\end{definition}
Thusly, then the G-operator in details is given by
\begin{align}
  {{\hat{H}}^{\left( gr \right)}}& ={{\hat{H}}^{\left( cl \right)}}+{{E}^{\left( \operatorname{Im} \right)}}={{\hat{H}}^{\left( cl \right)}}+\sqrt{-1}\hbar \hat{w} \notag\\
 & =\sqrt{-1}\hbar \left( {{\partial }_{t}}+\hat{w} \right) \notag\\
 & =\sqrt{-1}\hbar \hat{{{D}_{t}}} \notag
\end{align}
That T-operator is then expressed as ${{\hat{H}}^{\left( gr \right)}}/\hbar =\hat{{{N}_{t}}}$.  Thusly, the eigenvalue equation $\hat{{{N}_{t}}}\psi ={{N}_{t}}\psi$ follows.
As we stated, G-dynamics $\hat{w}$ is a Hermitian operator in the most of time, it implies that the
the G-operator is a non-Hermitian operator. As a result of this point, we say that the eigenvalue of the G-operator must be in a complex form.
\begin{theorem}
  The eigenvalue of G-operator is denoted as
 ${{E}^{\left( gr \right)}}=\hbar {{N}_{t}}$,
where the eigenvalue of T-operator is
\begin{align}
  {{N}_{t}}& ={{E}^{\left( gr \right)}}/\hbar ={{E}^{\left( cl \right)}}/\hbar +\sqrt{-1}{{E}^{\left( g \right)}}/\hbar={{E}^{\left( cl \right)}}/\hbar +\sqrt{-1}{{w}}\notag
\end{align}
where  ${{E}^{\left( g \right)}}=\hbar {{w}}$.
\begin{proof}
According to the definition \ref{d3} of G-operator, the eigenvalue equation can be more specifically expressed below
\begin{align}\label{eq8}
 {{\hat{H}}^{\left( gr \right)}}\psi & ={{\hat{H}}^{\left( cl \right)}}\psi +{{E}^{\left( \operatorname{Im} \right)}}\psi ={{\hat{H}}^{\left( cl \right)}}\psi +\sqrt{-1}\hbar \hat{w}\psi  \\
 & =\sqrt{-1}\hbar \left( {{\partial }_{t}}+\hat{w} \right)\psi =\sqrt{-1}\hbar \hat{{{D}_{t}}}\psi  =\sqrt{-1}\hbar {{D}_{t}}\psi  \notag \\
 & =\sqrt{-1}\hbar \left( {{\partial }_{t}}\psi +{{w}}\psi  \right) \notag \\
 & =\left( {{E}^{\left( cl \right)}}+\sqrt{-1}\hbar {{w}} \right)\psi  \notag \\
 & ={{E}^{\left( gr\right)}}\psi  \notag
\end{align}
Hence, we obtain the eigenvalue of the G-operator that is given by
${{E}^{\left( gr\right)}}=\sqrt{-1}\hbar {{D}_{t}}$,  it's rewritten as
${{E}^{\left( gr \right)}}={{E}^{\left( cl \right)}}+\sqrt{-1}{{E}^{\left( g \right)}}$,
and the eigenvalue of the D-operator $\hat{{{D}_{t}}}$ follows
\[{{D}_{t}}={{E}^{\left(gr \right)}}/\sqrt{-1}\hbar ={{E}^{\left( cl \right)}}/\sqrt{-1}\hbar +{{w}}\]
Similarly, the eigenvalue of the T-operator $\hat{{{N}_{t}}}$ also emerges as follows
${{N}_{t}}={{E}^{\left( gr \right)}}/\hbar$,
where
${{E}^{\left( g \right)}}/\hbar ={{w}}$, therefore,
the eigenvalue of the operator $\hat{{{N}_{t}}}$ can be rewritten as
\begin{align}
 {{N}_{t}} &={{E}^{\left( gr \right)}}/\hbar ={{E}^{\left( cl \right)}}/\hbar +\sqrt{-1}{{E}^{\left( g \right)}}/\hbar  \notag
\end{align}
Note that the eigenvalue of the G-operator can be rewritten in a form
${{E}^{\left(gr\right)}}=\hbar {{N}_{t}}$. Therefore, we complete the proof as desired.
\end{proof}
\end{theorem}
Notice that the eigenvalue of G-operator and T-operator are complex form, these are the features of the non-Hermitian operator.
As G-dynamics stated, it's generated by the structure function, and it forms the G-operator, it reveals that there exists a complete Schr\"{o}dinger equation such that the structure function is naturally involved in the equation.
To sum up, the three operators defined in this section are listed as follows:

D-operator: $\hat{{{D}_{t}}}={{\partial }_{t}}\pm\hat{w}$, it mainly studies $\hat{{{D}_{t}}}={{\partial }_{t}}+\hat{w}$.

T-operator: $\hat{{{N}_{t}}}=\sqrt{-1}\hat{{{D}_{t}}}=\sqrt{-1}{{{\partial}_{t}}}+\sqrt{-1}\hat{w}$.

G-operator: ${{\hat{H}}^{\left( gr \right)}}=\hbar \hat{{{N}_{t}}}=\sqrt{-1}\hbar\hat{{{D}_{t}}}$.

Above three operators are the extension in terms of the time derivative, in some ways.

\section{Geometrinetic energy operator (GEO) and Ri-operator}
As a result of the G-operator, we use geomentum operator \ref{d5}  to reconstruct the new Hamiltonian operator.  In order to smoothly build such complete new Hamiltonian operator, we firstly need to define a corresponding kinetic energy operator associated with the geometric
potential function $s$, therefore, let's rewrite
geomentum operator as ${{\hat{P}}^{\left( ri \right)}}=-\sqrt{-1}\hbar D$ from definition \ref{d5},  then we give the definition below based on the geomentum operator.

\begin{definition}[Geometrinetic energy operator (GEO)]
  Geometrinetic energy operator (GEO) is defined as
  \[{{\hat{T}}^{\left( ri \right)}}=\frac{{{\hat{P}}^{\left( ri \right)2}}}{2m}=-\frac{{{\hbar }^{2}}}{2m}\left( \Delta +2\nabla s\cdot \nabla \right)+{{T}^{\left( c \right)}}\]where ${{T}^{\left( c \right)}}=-\frac{{{\hbar }^{2}}}{2m}\left( \Delta s+\nabla s\cdot \nabla s \right)$ is called structural $c$-energy.

\end{definition}
More precisely, the derivation of the geometrinetic energy operator as an extension of the kinetic energy operator can be given by logically computation.
Since the square of generalized gradient operator that is calculated as
\[{{D}^{2}}\psi=\Delta \psi+2\nabla s\cdot \nabla \psi+\psi\left( \Delta s+\nabla s\cdot \nabla s \right)\]Accordingly, it yields an second order operator
\begin{equation}\label{eq7}
  {{D}^{2}}=\Delta+2\nabla s\cdot \nabla +\left( \Delta s+\nabla s\cdot \nabla s \right)
\end{equation}where $\Delta$ is the Laplacian.
Then the eigenvalue equation of the geometrinetic energy operator ${{\hat{T}}^{\left( ri \right)}}$ with respect to the wave function $\psi$ follows
\[{{\hat{T}}^{\left( ri \right)}}\psi=\frac{{{\hat{P}}^{\left( ri \right)2}}}{2m}\psi=-\frac{{{\hbar }^{2}}}{2m}\left( \Delta \psi+2\nabla s\cdot \nabla \psi \right)+\psi{{T}^{\left( c \right)}}\]

\begin{definition}\label{d7}
Ri-operator is defined as
 ${{\hat{H}}^{\left( ri \right)}}={{\hat{T}}^{\left( ri \right)}}+V$,
where ${{\hat{T}}^{\left( ri \right)}}$ is the geometrinetic energy operator.
\end{definition}
As a result, Ri-operator is a natural extension of the classical Hamiltonian operator \eqref{eq15}, then it can be rewritten in the form
\begin{equation}\label{eq14}
  {{\hat{H}}^{\left( ri \right)}}={{\hat{H}}^{\left( cl \right)}}-\frac{{{\hbar }^{2}}}{m}\nabla s\cdot \nabla +{{T}^{\left( c \right)}}
\end{equation}
by using the classical Hamiltonian operator \eqref{eq15}.

\subsection{The G-dynamics with Schr\"{o}dinger equation}
This section, we will mainly discuss the eigenvalue equation of G-dynamics based on the Schr\"{o}dinger equation \eqref{eq2}, it nicely proves that G-operator and T-operator are natural results on G-dynamics.
Rewriting the G-dynamics that is originally proposed as
\[\hat{w}=\frac{1}{\sqrt{-1}\hbar }{{\left[ s,\hat{H} \right]}_{QPB}}={{E}^{\left( \operatorname{Im} \right)}}/\sqrt{-1}\hbar\]
where the Hamiltonian operator $\hat{H}$ in it is completely determined by the physical truth.

Actually, when the Hamiltonian operator \eqref{eq4}  ${{\hat{H}}^{\left( cl \right)}}=\sqrt{-1}\hbar\frac{\partial}{\partial t}$ is considered to the formula of the G-dynamics ${{\hat{w}}^{\left( cl \right)}}$, it gets  ${{w}^{\left(q\right)}}=-{{\partial }_{t}}s$. More precisely,
\[{{\hat{w}}^{\left( cl \right)}}\psi =\frac{1}{\sqrt{-1}\hbar }{{\left[ s,{{\hat{H}}^{\left( cl \right)}} \right]}_{QPB}}\psi =\frac{1}{\sqrt{-1}\hbar }\left( s{{\hat{H}}^{\left( cl \right)}}\psi -{{\hat{H}}^{\left( cl \right)}}\left( s\psi  \right) \right)\]
Plugging the Schr\"{o}dinger equation \eqref{eq2} into above eigenvalue equation, we obtain
\[\sqrt{-1}\hbar {{\partial }_{t}}\psi ={{\hat{H}}^{\left( cl \right)}}\psi ,~~\sqrt{-1}\hbar {{\partial }_{t}}\left( s\psi  \right)={{\hat{H}}^{\left( cl \right)}}\left( s\psi  \right)\]
and the eigenvalue equation of G-dynamics ${{\hat{w}}^{\left( cl \right)}}$ as an operator is
\begin{align}\label{eq12}
 {{\hat{w}}^{\left( cl \right)}}\psi &=\frac{1}{\sqrt{-1}\hbar }{{\left[ s,{{\hat{H}}^{\left( cl \right)}} \right]}_{QPB}}\psi =\frac{1}{\sqrt{-1}\hbar }\left( s\sqrt{-1}\hbar {{\partial }_{t}}\psi -\sqrt{-1}\hbar {{\partial }_{t}}\left( s\psi  \right) \right) \notag\\
 & =\left( s{{\partial }_{t}}\psi -{{\partial }_{t}}\left( s\psi  \right) \right) \notag\\
 & =-\psi {{\partial }_{t}}s \notag
\end{align}If we denote ${{\hat{w}}^{\left( cl \right)}}\psi ={{w}^{\left(q \right)}}\psi$ here, then it certainly gets the specific eigenvalue for ${{\hat{w}}^{\left( cl \right)}}$  given by ${{w}^{\left( q\right)}}=-{{\partial }_{t}}s$.
Note that this has indicated a fact that for any function form given for
the G-dynamics $\hat{w}$ in such condition, the eigenvalue is always dependent on the structure function.

Based on the Schr\"{o}dinger equation \eqref{eq2}, it gets an equation similar to the heat equation that is given by
  \[{{\partial }_{t}}s =\frac{\sqrt{-1}\hbar }{m}\left( \Delta s/2+\nabla s\cdot \nabla \ln \psi  \right)\]
For this equation, we take the \eqref{eq15} into the formula of the G-dynamics $\hat{w}$, and it deduces a result
\begin{align}
 {{\hat{w}}^{\left( cl \right)}}\psi&=\frac{1}{\sqrt{-1}\hbar }{{\left[ s,{{\hat{H}}^{\left( cl \right)}} \right]}_{QPB}}\psi =\frac{1}{\sqrt{-1}\hbar }{{\left[ s,-\frac{{{\hbar }^{2}}}{2m}\Delta +V \right]}_{QPB}}\psi  \notag\\
 & =-\frac{\hbar }{2\sqrt{-1}m}{{\left[ s,\Delta  \right]}_{QPB}}\psi  \notag\\
 & =\frac{\hbar }{2\sqrt{-1}m}\left( 2\nabla s\cdot \nabla +\Delta s \right) \psi \notag
\end{align}Due to the Schr\"{o}dinger equation\eqref{eq2},  it yields
\[{{\hat{w}}^{\left( cl \right)}}\psi =-\psi {{\partial }_{t}}s=\frac{\hbar }{2\sqrt{-1}m}\left( \Delta s+2\nabla s\cdot \nabla  \right)\psi \]where $-{{\hat{w}}^{\left( cl \right)}}\psi =\psi {{\partial }_{t}}s$ has used.
It's definite to form an equation similar to heat equation associated with the geometric potential function
\begin{align}
 {{\partial }_{t}}s & =\frac{\sqrt{-1}\hbar }{2m\psi }\left( \psi \Delta s+2\nabla s\cdot \nabla \psi  \right)=\frac{\sqrt{-1}\hbar }{m}\left( \Delta s/2+\nabla s\cdot \nabla \ln \psi  \right) \notag
\end{align}

The G-dynamics in terms of the Hamiltonian operator \eqref{eq15} in terms of coordinates can be expressed as
\begin{equation}\label{eq17}
  {{\hat{w}}^{\left( cl \right)}}=\frac{\hbar }{2\sqrt{-1}m}\left( 2\nabla s\cdot \nabla +\Delta s \right)=\frac{\hbar }{2\sqrt{-1}m}\hat{Q}
\end{equation}It can be called the G-dynamics of type I, geometric differential operator $\hat{Q}= 2\nabla s\cdot \nabla +\Delta s$ is the curvature
operator.
As we can see, the G-dynamics on such expression given by the \eqref{eq17} is totally determined by the  geometric potential function.  Geometrinetic energy operator by using the G-dynamics ${{\hat{w}}^{\left( cl \right)}}$ can be rewritten as \[{{\hat{T}}^{\left( ri \right)}}={{\hat{T}}^{\left( cl \right)}}-{E}^{\left( s \right)}/2-\sqrt{-1}\hbar {{\hat{w}}^{\left( cl \right)}}\]
where

The imaginary geomenergy:  ${{E}^{\left( \operatorname{Im} \right)}}\left( {{{\hat{w}}}^{\left( cl \right)}} \right)=\sqrt{-1}\hbar {{\hat{w}}^{\left( cl \right)}}$,

Classical kinetic energy operator: ${{\hat{T}}^{\left( cl \right)}}=-\frac{{{\hbar }^{2}}}{2m}\Delta$,

The geometric potential energy function: ${{E}^{\left( s \right)}}=\frac{{{\hbar }^{2}}}{m}\nabla s\cdot \nabla s$.

As a consequence, we can get an equation similar to Schr\"{o}dinger equation as follows
$$ \sqrt{-1}\hbar {{\hat{w}}^{\left( cl \right)}}\psi =\frac{{{\hbar }^{2}}}{2m}
   \left( 2\nabla s\cdot \nabla \psi+\psi\Delta s  \right) $$
Therefore, it leads to a parallel system of equations
\[\left\{ \begin{matrix}
 \sqrt{-1}\hbar {{\hat{w}}^{\left( cl \right)}}\psi =\frac{{{\hbar }^{2}}}{2m}
   \left( 2\nabla s\cdot \nabla \psi+\psi\Delta s\right)  \\
   \sqrt{-1}\hbar {{\partial }_{t}}\psi =-\frac{{{\hbar }^{2}}}{2m}\Delta \psi +V\psi   \\
\end{matrix} \right.\]
for wave function $\psi$. It's convinced that the anti-Hermitian equation induced by the G-dynamics of type I is a supplement to Schr\"{o}dinger equation, they exist side by side and play a part together.

Take a deep analysis to above two equations, if let $-{{\hat{w}}^{\left( cl \right)}}\psi =\psi {{\partial }_{t}}s$ be given for substitution,  obviously, it yields
\[\sqrt{-1}\hbar \psi {{\partial }_{t}}s=-\frac{{{\hbar }^{2}}}{2m}\left( \psi \Delta s+2\nabla s\cdot \nabla \psi  \right)\]
It follows the bonding equation
\[\sqrt{-1}\hbar \left( {{\partial }_{t}}\psi +\psi {{\partial }_{t}}s \right)=-\frac{{{\hbar }^{2}}}{2m}\left( \Delta \psi +2\nabla s\cdot \nabla \psi +\psi \Delta s \right)+V\psi \]This is a natural generalization of classical Schr\"{o}dinger equation, it helps us to think quantum problems more wider and their explanations.
Definitely, it can be rewritten as
\begin{align}
 \sqrt{-1}\hbar \left( {{\partial }_{t}}+{{\partial }_{t}}s \right)\psi  &=-\frac{{{\hbar }^{2}}}{2m}\left( \Delta +2\nabla s\cdot \nabla +\Delta s+\nabla s\cdot \nabla s \right)\psi  \notag\\
 & \begin{matrix}
   {} & {} & {} & {} & {} & {} & {}& {} & {} & {} \\
\end{matrix}+\left( V+\frac{{{\hbar }^{2}}}{2m}\nabla s\cdot \nabla s \right)\psi \notag \\
 & ={{\hat{H}}^{\left( ri \right)}}\psi+\frac{{{\hbar }^{2}}}{2m}\psi \nabla s\cdot \nabla s \notag
\end{align}
Or in the form
\begin{equation}\label{eq18}
  \sqrt{-1}\hbar \left( {{\partial }_{t}}-{{\hat{w}}^{\left( cl \right)}} \right)\psi ={{\hat{H}}^{\left( ri \right)}}\psi+{{E}^{\left( s \right)}}\psi/2
\end{equation}It evidently implies that ${{E}^{\left( s \right)}}$ plays the role of potential energy.  In fact, the \eqref{eq18} supports a statement to the definition \ref{d3}. In other words, such definition \ref{d3} is a natural extension for the classical results, in the later discussions, we can surely say about this points.

\begin{definition}\label{d8}
 D-operator based on the Schr\"{o}dinger equation \eqref{eq2} is
$\hat{{{D}_{t}}}^{\left( cl \right)}={{\partial }_{t}}-{{\hat{w}}^{\left( cl \right)}}$, and then T-operator is  $${{\hat{N}}_{t}}^{\left( cl \right)}=\sqrt{-1}\hat{{{D}_{t}}}^{\left( cl \right)}=\sqrt{-1}{{{\partial}_{t}}}-\sqrt{-1}\hat{w}^{\left( cl \right)}$$ and
 the G-operator in terms of the G-dynamics of type I is given by \[{{\hat{H}}^{\left(gr \right)}}=\hbar {{\hat{N}}_{t}}^{\left( cl \right)}={{\hat{H}}^{\left( cl \right)}}-\sqrt{-1}\hbar {{\hat{w}}^{\left( cl \right)}}\] where here ${{\hat{H}}^{\left( cl \right)}}=\sqrt{-1}\hbar{{{\partial}_{t}}}$, and  the imaginary geomenergy is ${{E}^{\left( \operatorname{Im} \right)}}\left( {{{\hat{w}}}^{\left( cl \right)}} \right)=\sqrt{-1}\hbar {{\hat{w}}^{\left( cl \right)}}$.
\end{definition}
As a consequence of the definition \ref{d3}, we figure out that eigenvalue theorem below.
\begin{theorem}\label{t3}
 The eigenvalue of the G-operator in definition \ref{d8} is
  \[{{E}^{\left( gr \right)}}={{E}^{\left( cl \right)}}-\sqrt{-1}\hbar {{w}^{\left(q\right)}}\]
 \begin{proof}
 For a given wave function $\psi$, the eigenvalue equation of the G-operator on Schr\"{o}dinger equation \eqref{eq2} is expressed as
\begin{align}
{{\hat{H}}^{\left( gr \right)}}\psi  & =\hbar {{\hat{N}}_{t}}^{\left( cl \right)}\psi ={{\hat{H}}^{\left( cl \right)}}\psi -{{E}^{\left( \operatorname{Im} \right)}}\left( {{{\hat{w}}}^{\left( cl \right)}} \right)\psi  \notag\\
 & ={{\hat{H}}^{\left( cl \right)}}\psi -\sqrt{-1}\hbar {{{\hat{w}}}^{\left( cl \right)}}\psi  \notag\\
 & ={{E}^{\left( cl \right)}}\psi -\sqrt{-1}\hbar \psi {{w}^{\left(q\right)}} \notag\\
 & =\left( {{E}^{\left( cl \right)}}-\sqrt{-1}\hbar {{w}^{\left(q\right)}} \right)\psi  \notag\\
 & ={{E}^{\left( gr \right)}}\psi  \notag
\end{align}where ${{\hat{w}}^{\left( cl \right)}}\psi ={{w}^{\left( q \right)}}\psi $ has been used for the proof.
Then, we complete the proof.
 \end{proof}
\end{theorem}
In fact, due to the conjugate property of complex numbers, then it also has
${{E}^{\left(gr \right)}}={{E}^{\left( cl \right)}}+\sqrt{-1}\hbar {{w}^{\left(q\right)}}$ with respect to the corresponding eigenfunctions.

\section{The cases of G-dynamics of three types}

\subsection{For classical Hamiltonian operator}

To consider quantum harmonic oscillator \eqref{eq3},  as the G-dynamics of it shown by \eqref{eq6},
\[{{\hat{w}}^{\left( cl \right)}}=\frac{\sqrt{-1}}{\hbar }\frac{{{\hat{p}}^{\left( cl \right)2}}s}{2m}+\frac{\sqrt{-1}}{m\hbar }{{\hat{p}}^{\left( cl \right)}}s{{\hat{p}}^{\left( cl \right)}}=b_{c}\left( 2\nabla s\cdot \nabla +\Delta s \right)\]
The pure imaginary geomenergy form is
\begin{align}
 {{E}^{\left( \operatorname{Im} \right)}}\left( {{\hat{w}}^{\left( cl \right)}} \right) & =\sqrt{-1}\hbar {{\hat{w}}^{\left( cl \right)}} =-\frac{{{\hat{p}}^{\left( cl \right)2}}s}{2m}-\frac{1}{m}{{\hat{p}}^{\left( cl \right)}}s{{\hat{p}}^{\left( cl \right)}} \notag\\
 & =\frac{{{\hbar }^{2}}}{2m}
   \left( 2\nabla s\cdot \nabla +\Delta s  \right) \notag
\end{align}
Let's see how its characteristic equation shows, it's given by
\begin{align}
 {{\hat{w}}^{\left( cl \right)}}\psi & =\frac{\sqrt{-1}}{\hbar }\psi \frac{{{\hat{p}}^{\left( cl \right)2}}s}{2m}+\frac{\sqrt{-1}}{m\hbar }{{\hat{p}}^{\left( cl \right)}}s{{\hat{p}}^{\left( cl \right)}}\psi  \notag\\
 & =-\frac{\sqrt{-1}\hbar }{2m}\left( \psi \Delta s+2\nabla s\cdot \nabla \psi  \right) \notag\\
 & =\frac{\hbar }{2\sqrt{-1}m}\left( \Delta s+2\nabla s\cdot \nabla  \right)\psi  \notag
\end{align}where $\Delta ={{\nabla }^{2}}$ is the Laplacian, then it gets
$$\sqrt{-1}\hbar {{\hat{w}}^{\left( cl \right)}}\psi =\frac{{{\hbar }^{2}}}{2m}
   \left( 2\nabla s\cdot \nabla\psi +\psi\Delta s  \right)$$
Subsequently, it derives the G-dynamics \eqref{eq17},
or in the form
$$\frac{2\sqrt{-1}m}{\hbar }{{\hat{w}}^{\left( cl \right)}}=\Delta s+2\nabla s\cdot \nabla$$
by using \eqref{eq7}, we get its another expression in terms of the G-dynamics shown as
\[{{D}^{2}}=\frac{2\sqrt{-1}m}{\hbar }{{\hat{w}}^{\left( cl \right)}}+\Delta +\nabla s\cdot \nabla s\]
Geometrinetic energy operator (GEO) can be rewritten in a form
\begin{align}
 {{\hat{T}}^{\left( ri \right)}} &=-\frac{{{\hbar }^{2}}}{2m}{{D}^{2}}=-\frac{{{\hbar }^{2}}}{2m}\left( \frac{2\sqrt{-1}m}{\hbar }{{\hat{w}}^{\left( cl \right)}}+\Delta +\nabla s\cdot \nabla s \right) \notag\\
 & =-\frac{{{\hbar }^{2}}}{2m}\Delta -\frac{{{\hbar }^{2}}}{2m}\nabla s\cdot \nabla s-\sqrt{-1}\hbar {{\hat{w}}^{\left( cl \right)}} \notag\\
 & =-\frac{{{\hbar }^{2}}}{2m}\left( \Delta +\nabla s\cdot \nabla s \right)-\sqrt{-1}\hbar {{\hat{w}}^{\left( cl \right)}} \notag\\
 & =-\frac{{{\hbar }^{2}}}{2m}\Delta -\frac{{{E}^{\left( s \right)}}}{2}-\sqrt{-1}\hbar {{\hat{w}}^{\left( cl \right)}} \notag
\end{align}
Therefore, the Ri-operator is given by
$${{\hat{H}}^{\left( ri \right)}}={{\hat{T}}^{\left( ri \right)}}+V={{\hat{H}}^{\left( cl \right)}}-\frac{{{E}^{\left( s \right)}}}{2}-\sqrt{-1}\hbar {{\hat{w}}^{\left( cl \right)}}$$
Consider the one-dimensional quantum harmonic oscillator \eqref{eq3}, then the Ri-operator has a specific expression given by
$${{\hat{H}}^{\left( ri \right)}}={\frac  {{{\hat{p}}}^{{\left( cl \right)2}}}{2m}}+{\frac  {m\omega ^{{2}}x^{{2}}}{2}}-\frac{{{E}^{\left( s \right)}}}{2}-\sqrt{-1}\hbar {{\hat{w}}^{\left( cl \right)}}$$
This helps to study more general quantum harmonic oscillator \eqref{eq3}.

\subsection{For Ri-operator}

\begin{theorem}\label{t4}
 The G-dynamics in terms of ${{\hat{H}}^{\left( ri \right)}}$ is
  $$\hat{w}^{\left( ri \right)}={{\hat{w}}^{\left( cl \right)}}+w^{\left(s\right)}$$denoted as the G-dynamics of type II,  and ${{\hat{w}}^{\left( cl \right)}}$ is the G-dynamics of type I,  $w^{\left(s\right)}=\frac{\hbar }{\sqrt{-1}m}\nabla s\cdot \nabla s$ is the G-dynamics of type III.
\begin{proof}
For this case, let's consider ${{\hat{H}}^{\left( ri \right)}}={{\hat{T}}^{\left( ri \right)}}+V$ given by Ri-operator \ref{d7}, then the G-dynamics can be calculated as
\[\hat{w}^{\left( ri \right)}=\frac{1}{\sqrt{-1}\hbar }{{\left[ s,{{\hat{H}}^{\left( ri \right)}} \right]}_{QPB}}=\frac{1}{\sqrt{-1}\hbar }{{\left[ s,{{\hat{T}}^{\left( ri \right)}}+V \right]}_{QPB}}\]where $V$ is a real potential function.  Thusly, we obtain
\begin{align}
 {{\left[ s,{{\hat{T}}^{\left( ri \right)}} \right]}_{QPB}}\psi & =-\frac{{{\hbar }^{2}}}{2m}{{\left[ s,\Delta +2\nabla s\cdot \nabla  \right]}_{QPB}}\psi \notag \\
 & =-\frac{{{\hbar }^{2}}}{2m}{{\left[ s,\Delta  \right]}_{QPB}}\psi -\frac{{{\hbar }^{2}}}{m}{{\left[ s,\nabla s\cdot \nabla  \right]}_{QPB}}\psi  \notag
\end{align}
where
\begin{align}
 {{\left[ s,\Delta  \right]}_{QPB}}\psi & =s\Delta \psi -\Delta \left( s\psi  \right) \notag\\
 & =-2\nabla s\cdot \nabla \psi -\psi \Delta s \notag\\
 & =-\left( 2\nabla s\cdot \nabla +\Delta s \right)\psi \notag
\end{align}and
\[{{\left[ s,\nabla s\cdot \nabla  \right]}_{QPB}}\psi =-\psi \nabla s\cdot \nabla s\]Therefore, we get
\[{{\left[ s,{{\hat{T}}^{\left( ri \right)}} \right]}_{QPB}}=\frac{{{\hbar }^{2}}}{2m}\left( 2\nabla s\cdot \nabla +\Delta s+2\nabla s\cdot \nabla s \right)\]Thusly,
\begin{align}\label{eq22}
 \hat{w}^{\left( ri \right)}&=\frac{1}{\sqrt{-1}\hbar }{{\left[ s,{{\hat{T}}^{\left( ri \right)}} \right]}_{QPB}}=\frac{\hbar }{2\sqrt{-1}m}\left( 2\nabla s\cdot \nabla +\Delta s+2\nabla s\cdot \nabla s \right) \\
 & =\frac{\hbar }{2\sqrt{-1}m}\left( 2\nabla s\cdot \nabla +\Delta s \right)+\frac{\hbar }{\sqrt{-1}m}\nabla s\cdot \nabla s \notag\\
 & ={{\hat{w}}^{\left( cl \right)}}+w^{\left(s\right)}\notag
\end{align}
where  $w^{\left(s\right)}={{\left[ {{\hat{w}}^{\left( cl \right)}},s \right]}_{QPB}}=\frac{\hbar }{\sqrt{-1}m}\nabla s\cdot \nabla s$ is denoted.
\end{proof}
\end{theorem}
Therefore, there has three different kinds of G-dynamics as a quantum operator found as follows:
\begin{description}
  \item[I:] ${{{\hat{w}}}^{\left( cl \right)}}=\frac{\hbar }{2\sqrt{-1}m}\left( 2\nabla s\cdot \nabla+\Delta s  \right)  $.
  \item[II:] ${{{\hat{w}}}^{\left( ri \right)}}=\frac{\hbar }{2\sqrt{-1}m}\left( 2\nabla s\cdot \nabla +\Delta s+2\nabla s\cdot \nabla s \right)$.

  \item[III:] ${{w}^{\left( s \right)}}=\frac{\hbar }{\sqrt{-1}m}\nabla s\cdot \nabla s$.
\end{description}
where $\nabla s$ can be realized as line curvature which is the gradient of the geometric potential function $s$, these three different types can be found so far, it needs more studies on them.
Obviously, ${{\hat{w}}^{\left( ri \right)}}-{{\hat{w}}^{\left( cl \right)}}={{w}^{\left( s \right)}}$ holds.  The theorem \ref{t4} can be rewritten in a form
\[{{E}^{\left( \operatorname{Im} \right)}}\left( {\hat{w}^{\left( ri \right)}} \right)=\sqrt{-1}\hbar \hat{w}^{\left( ri \right)}=\sqrt{-1}\hbar {{\hat{w}}^{\left( cl \right)}}+{{E}^{\left( s \right)}}\]where  ${{E}^{\left( s \right)}}=\frac{{{\hbar }^{2}}}{m}\nabla s\cdot \nabla s=\sqrt{-1}\hbar {{w}^{\left( s \right)}}$,
or it can be expressed by the imaginary geomenergy
\[{{E}^{\left( \operatorname{Im} \right)}}\left( {\hat{w}^{\left( ri \right)}} \right)={{E}^{\left( \operatorname{Im} \right)}}\left( {{\hat{w}}^{\left( cl \right)}} \right)+{{E}^{\left( s \right)}}\]
where ${{E}^{\left( \operatorname{Im} \right)}}\left( {{\hat{w}}^{\left( cl \right)}} \right)=\sqrt{-1}\hbar {{\hat{w}}^{\left( cl \right)}}$ is  the imaginary geomenergy in terms of ${{\hat{w}}^{\left( cl \right)}}$.
Correspondingly, the structural $c$-energy can be given in a form
\begin{align}
{{T}^{\left( c \right)}}  & =-\frac{{{\hbar }^{2}}}{2m}\left( \Delta s+{{\left| \nabla s \right|}^{2}} \right)=-\frac{{{\hbar }^{2}}}{2m}\Delta s-\frac{{{\hbar }^{2}}}{2m}{{\left| \nabla s \right|}^{2}}\notag \\
 & =-\frac{{{\hbar }^{2}}}{2m}\Delta s-{{E}^{\left( s \right)}}/2 \notag
\end{align}
where it results in ${{T}^{\left( c \right)}}+{{E}^{\left( s \right)}}/2=-\frac{{{\hbar }^{2}}}{2m}\Delta s$.

There is more we need to say clearly,
\[{{\hat{w}}^{\left( ri \right)}}={{\hat{w}}^{\left( cl \right)}}+{{w}^{\left( s \right)}}={{\hat{w}}^{\left( cl \right)}}-\sqrt{-1}a_{c}\nabla s\cdot \nabla s\]
where $a_{c}=\hbar /m$.
Then the conjugate transpose of the ${{\hat{w}}^{\left( ri \right)}}$ is given by \[{{\hat{w}}^{\left( ri \right)\dagger }}={{\hat{w}}^{\left( cl \right)\dagger }}+\sqrt{-1}a_{c}\nabla s\cdot \nabla s\]Accordingly, it yields ${{\hat{w}}^{\left( ri \right)}}+{{\hat{w}}^{\left( ri \right)\dagger }}=2{{\hat{w}}^{\left( cl \right)}}$. Meanwhile, let ${{\hat{w}}^{\left( g \right)}}$ be denoted as NG operator to fit
\[{{\hat{w}}^{\left( g \right)}}={{\hat{w}}^{\left( ri \right)}}-{{\hat{w}}^{\left( ri \right)\dagger }}=-2\sqrt{-1}a_{c}\nabla s\cdot \nabla s\ne 0\]
Obviously, this NG operator breaks the Hermiticity as usually understood.

\subsection{G-operator for Ri-operator}

To analyze G-dynamics for both situations together, we can obtain some useful results.
\begin{align}
  \hat{w}^{\left( ri \right)}\psi& ={{\hat{w}}^{\left( cl \right)}}\psi +w^{\left(s\right)}\psi \notag\\
 & =\frac{\hbar }{2\sqrt{-1}m}\left( \Delta s+2\nabla s\cdot \nabla  \right)\psi +\frac{\hbar }{\sqrt{-1}m}\psi \nabla s\cdot \nabla s \notag
\end{align}
Accordingly, it leads to the equation of imaginary geomenergy
\[\sqrt{-1}\hbar \hat{w}^{\left( ri \right)}\psi =\frac{{{\hbar }^{2}}}{2m}\left( 2\nabla s\cdot \nabla +\Delta s+2\nabla s\cdot \nabla s \right)\psi \]
and with the Schr\"{o}dinger equation \eqref{eq2}, it generates
\begin{align}
\sqrt{-1}\hbar {{\partial }_{t}}\psi -\sqrt{-1}\hbar \hat{w}^{\left( ri \right)}\psi &=\sqrt{-1}\hbar \left( {{\partial }_{t}}-\hat{w}^{\left( ri \right)} \right)\psi  \notag\\
 & =-\frac{{{\hbar }^{2}}}{2m}\Delta \psi +V\psi -\frac{{{\hbar }^{2}}}{2m}\left( 2\nabla s\cdot \nabla +\Delta s+2\nabla s\cdot \nabla s \right)\psi  \notag\\
 & =-\frac{{{\hbar }^{2}}}{2m}\left( \Delta +2\nabla s\cdot \nabla +\Delta s+\nabla s\cdot \nabla s \right)\psi +\left( V-{{E}^{\left( s \right)}}/2\right)\psi  \notag\\
 &={{\hat{H}}^{\left( ri \right)}}\psi-{{E}^{\left( s \right)}}\psi/2 \notag
\end{align}
As a result of equation \eqref{eq18} together, we get a similar equations given by
\begin{align}\label{eq19}
  & \sqrt{-1}\hbar \left( {{\partial }_{t}}-{{\hat{w}}^{\left( cl \right)}} \right)\psi ={{{\hat{H}}}^{\left( ri \right)}}\psi+{{E}^{\left( s \right)}}\psi/2  \\
 & \sqrt{-1}\hbar \left( {{\partial }_{t}}-{{\hat{w}}^{\left( ri \right)}} \right)\psi ={{{\hat{H}}}^{\left( ri \right)}}\psi-{{E}^{\left( s \right)}}\psi/2 \notag
\end{align}
And then it rightly reflects the theorem \ref{t4}.

Ri-operator deduced by the G-operator is
\begin{equation}\label{a1}
  {{\hat{H}}^{\left( ri \right)}}={{\hat{H}}^{\left( gr \right)}}+{{E}^{\left( s \right)}}/2
\end{equation}
Since the \eqref{eq14} and theorem \ref{t4} hold, then
\begin{align}
 {{{\hat{H}}}^{\left( ri \right)}} &={{{\hat{H}}}^{\left( cl \right)}}-\frac{{{\hbar }^{2}}}{m}\nabla s\cdot \nabla +{{{{T}}}^{\left( c \right)}} \notag\\
 & ={{{\hat{H}}}^{\left( cl \right)}}-\sqrt{-1}\hbar {{\hat{w}}^{\left( ri \right)}}+{{E}^{\left( s \right)}}/2 \notag
\end{align}With the definition \ref{d3} of G-operator, it has
\[{{\hat{H}}^{\left( gr \right)}}={{\hat{H}}^{\left( cl \right)}}-\sqrt{-1}\hbar {{\hat{w}}^{\left( ri \right)}}=\sqrt{-1}\hbar {{\hat{D}}_{t}}^{\left(ri \right)}\]in terms of  the G-dynamics of type II,
where D-operator here is ${{\hat{D}}_{t}}^{\left( ri \right)}={{\partial }_{t}}-{{\hat{w}}^{\left( ri \right)}}$.
Conversely, note that the G-operator in terms of the G-dynamics of II is also expressed as
\begin{equation}\label{eq20}
  {{\hat{H}}^{\left( gr \right)}}={{\hat{H}}^{\left( ri \right)}}-{{E}^{\left( s \right)}}/2
\end{equation}
By using Ri-operator \eqref{a1}, \eqref{eq19} can be simplified as
\begin{align}\label{eq21}
  & \sqrt{-1}\hbar \left( {{\partial }_{t}}-{{{\hat{w}}}^{\left( cl \right)}} \right)\psi ={{{\hat{H}}}^{\left( gr \right)}}\psi +{{E}^{\left( s \right)}}\psi   \\
 & \sqrt{-1}\hbar \left( {{\partial }_{t}}-{{{\hat{w}}}^{\left( ri \right)}} \right)\psi ={{{\hat{H}}}^{\left( gr \right)}}\psi  \notag
\end{align}where G-operator in terms of the Ri-operator is definitely chosen as \[{{\hat{H}}^{\left( gr \right)}}=\sqrt{-1}\hbar \left( {{\partial }_{t}}-{{{\hat{w}}}^{\left( ri \right)}} \right)=\sqrt{-1}\hbar {{\hat{D}}_{t}}^{\left( ri \right)}\]and ${{\hat{D}}_{t}}^{\left( ri\right)}={{\partial }_{t}}-{{\hat{w}}^{\left( ri \right)}}$.
As a matter of fact, these two equations are equivalent to each other based on the theorem \ref{t4}.

As a consequence of the G-dynamics of type I, then it helps to construct four different kinds of the non-Hermitian Hamiltonian operators: Ri-operator $${{\hat{H}}^{\left( ri \right)}}={{{\hat{H}}}^{\left( cl \right)}}+{{{\hat{H}}}^{\left( s \right)}}={{{\hat{E}}}^{\left( w \right)}}+V-{{E}^{\left( s \right)}}/2={{{\hat{H}}}^{\left( cl \right)}}-{{E}^{\left( s \right)}}/2-\sqrt{-1}\hbar {{\hat{w}}^{\left( cl \right)}}$$The motor operator ${{\hat{E}}^{\left( w \right)}}={{c}_{1}}\Delta-\sqrt{-1}\hbar {{\hat{w}}^{\left( cl \right)}}$, the geometric Hamiltonian operator ${{\hat{H}}^{\left( s \right)}}=-{{E}^{\left( s \right)}}/2-\sqrt{-1}\hbar {{\hat{w}}^{\left( cl \right)}}$, and the G-operator $${{{\hat{H}}}^{\left( gr \right)}}={{{\hat{H}}}^{\left( cl \right)}}-{{E}^{\left( s \right)}}-\sqrt{-1}\hbar {{{\hat{w}}}^{\left( cl \right)}}$$ in terms of the G-dynamics of type II.

\subsection{Generalized covariant wave equation}
Correspondingly, the definition of generalized covariant wave equation as a revision of classic Schr\"{o}dinger equation follows based on geometrinetic energy operator.

The generalized covariant wave equation is given by \[\sqrt{-1}\hbar\hat{{{D}_{t}}}^{\left(ri \right)}\psi=\hbar \hat{{{N}_{t}}}^{\left(ri \right)}\psi={{{\hat{H}}}^{\left( gr \right)}}\psi\]by using \eqref{eq20},
where the D-operator is taken as $\hat{{{D}_{t}}}^{\left(ri \right)}={{\partial }_{t}}-{{{\hat{w}}}^{\left( ri \right)}}$.

Apparently, the generalized covariant wave equation can be given by employing \eqref{eq21},
\[\sqrt{-1}\hbar \left( {{\partial }_{t}}-{{{\hat{w}}}^{\left( cl \right)}}-{{w}^{\left( s \right)}} \right)\psi ={{\hat{H}}^{\left( gr \right)}}\psi \]
where $${{\hat{w}}^{\left( ri \right)}}={{\hat{w}}^{\left( cl \right)}}+{{E}^{\left( s \right)}}/\sqrt{-1}\hbar ={{\hat{w}}^{\left( cl \right)}}+{{w}^{\left( s \right)}}$$ is distinct to be seen from the theorem \ref{t4}, where ${{w}^{\left( s \right)}}={{E}^{\left( s \right)}}/\sqrt{-1}\hbar$.
More precisely about the generalized covariant wave equation are listed
\begin{align}
  & {{{\hat{H}}}^{\left( gr \right)}}=-\frac{{{\hbar }^{2}}}{2m}\left( \Delta +2\nabla s\cdot \nabla +\Delta s+2\nabla s\cdot \nabla s \right)={{{\hat{H}}}^{\left( ri \right)}}-{{E}^{\left( s \right)}}/2  \notag\\
 & {{{\hat{w}}}^{\left( ri \right)}}=b_{c}\left( 2\nabla s\cdot \nabla +\Delta s+2\nabla s\cdot \nabla s \right)={{{\hat{w}}}^{\left( cl \right)}}+{{w}^{\left( s \right)}}  \notag
\end{align}
Obviously, the generalized covariant wave equation is a natural complete extension of the classic Schr\"{o}dinger equation.

\section{GEO for generalized quantum harmonic oscillator}

The covariant time evolution of operators $x(t),{{\hat{p}}}^{{\left(ri \right)}}(t)$ depends on the Hamiltonian of the system. Considering the one-dimensional generalized quantum harmonic oscillator based upon geomentum operator \ref{d5}, it's given by
\begin{equation}\label{eq9}
  \hat{H}^{\left( ri \right)}={\frac  {{{\hat{p}}}^{{\left(ri \right)2}}}{2m}}+{\frac  {m\omega ^{{2}}x^{{2}}}{2}}={{\hat{H}}}^{{\left(cl \right)}}-\frac{{{E}^{\left( s \right)}}}{2}-\sqrt{-1}\hbar {{\hat{w}}^{\left( cl \right)}}
\end{equation}
where the classical Hamiltonian ${{\hat{H}}}^{{\left(cl \right)}}={\frac  {{{\hat{p}}}^{{\left( cl \right)2}}}{2m}}+{\frac  {m\omega ^{{2}}x^{{2}}}{2}}$ is given by \eqref{eq3},~ $V\left( x \right)=\frac{1}{2}m{{\omega }^{2}}{{x}^{2}}$,  the geomentum operator in one-dimensional is ${{\hat{p}}^{\left( ri \right)}}=-\sqrt{-1}\hbar \frac{\text{D}}{dx}$, and $\frac{\text{D}}{dx}=d/dx+u$. Accordingly,
the covariant evolution of the position and  geomentum operator based on the QCHS given by definition \ref{d10} are respectively given by:
\begin{equation}\label{eq10}
  {\mathcal{D}\over dt}x(t)={\sqrt{-1} \over \hbar }[\hat{H}^{\left(ri \right)},x(t)]
\end{equation}
\[{\mathcal{D}\over dt}{{\hat{p}}}^{{\left( ri\right)}}(t)={\sqrt{-1}\over \hbar }[\hat{H}^{\left( ri\right)},{{\hat{p}}}^{{\left( ri \right)}}(t)]\]where $[\cdot,\cdot]$ denotes the GGC of two operators.
More precisely, the concrete computational process of the first covariant evolution terms of the position is shown as
\[\left[ {{\hat{H}}^{\left( ri \right)}},x \right]={{\left[ {{\hat{H}}^{\left( ri \right)}},x \right]}_{QPB}}+G\left( s,{{\hat{H}}^{\left( ri \right)}},x \right)\]
Hence, the direct computation leads to a series of results
\begin{align}
 {{\left[ {{\hat{H}}^{\left( ri \right)}},x \right]}_{QPB}} & ={{\left[ \frac{{{\hat{p}}^{\left( ri \right)2}}}{2m},x \right]}_{QPB}}=\frac{1}{2m}{{\left[ {{\hat{p}}^{\left( ri \right)2}},x \right]}_{QPB}} \notag\\
 & =\frac{-{{\hbar }^{2}}}{2m}{{\left[ \frac{{{d}^{2}}}{d{{x}^{2}}}+2u\frac{d}{dx},x \right]}_{QPB}}\notag \\
 & =\frac{-{{\hbar }^{2}}}{m}\frac{\text{D}}{dx} \notag
\end{align}where \[{{\hat{p}}^{\left( ri \right)2}}=-{{\hbar }^{2}}\left( \frac{{{d}^{2}}}{d{{x}^{2}}}+2u\frac{d}{dx}+u_{x}+{u^{2}} \right)\]
and the QGB in terms of ${{\hat{H}}^{\left( ri \right)}},x$ is
\begin{align}
G\left( s,{{\hat{H}}^{\left( ri \right)}},x \right)  &=-x{{\left[ s,{{\hat{H}}^{\left( ri \right)}} \right]}_{QPB}}=x{{\left[ {{\hat{H}}^{\left( ri \right)}},s \right]}_{QPB}}\notag \\
 & =-\frac{{{\hbar }^{2}}}{m}xu\frac{d}{dx}-\frac{{{\hbar }^{2}}}{2m}x\left( u_{x}+2{{u}^{2}} \right) \notag
\end{align}
Then the QCPB about ${{\hat{H}}^{\left( ri \right)}},x$  specifically shows
\begin{align}
 \left[ {{\hat{H}}^{\left( ri \right)}},x \right] &={{\left[ {{\hat{H}}^{\left( ri \right)}},x \right]}_{QPB}}+G\left( s,{{\hat{H}}^{\left( ri \right)}},x \right) \notag\\
 & =\frac{-{{\hbar }^{2}}}{m}\frac{\text{D}}{dx}-\frac{{{\hbar }^{2}}}{m}xu\frac{d}{dx}-\frac{{{\hbar }^{2}}}{2m}x\left( u_{x}+2{{u}^{2}} \right) \notag\\
 & =-\frac{{{\hbar }^{2}}}{m}\left( \frac{\text{D}}{dx}+xu\frac{d}{dx}+\frac{1}{2}x\left( u_{x}+2{{u}^{2}} \right) \right) \notag
\end{align}
As a result, the covariant evolution of the position is given by
\begin{align}
  \frac{\mathcal{D}}{dt}x\left( t \right)& =\frac{\sqrt{-1}}{\hbar }\left[ {{\hat{H}}^{\left( ri \right)}},x \right] \notag\\
 & =-\frac{\sqrt{-1}\hbar }{m}\left( \frac{\text{D}}{dx}+xu\frac{d}{dx}+\frac{1}{2}x\left( u_{x}+2{{u}^{2}} \right) \right) \notag
\end{align}
In other words, the covariant evolution of the position based on the QCHS is
\[\frac{\mathcal{D}}{dt}x\left( t \right)=\frac{\sqrt{-1}}{\hbar }\left[ {{\hat{H}}^{\left( ri \right)}},x \right]=\frac{{{\hat{p}}^{\left( ri \right)}}}{m}+x\hat{w}^{\left( ri \right)}\]And the generalized Heisenberg equation with respect to $x$ follows
\begin{align}
 \frac{d}{dt}x\left( t \right) & =\frac{\sqrt{-1}}{\hbar }{{\left[ \hat{H}^{\left( ri \right)},x\left( t \right) \right]}_{QPB}}=\frac{{{\hat{p}}^{\left( ri \right)}}\left( t \right)}{m}\notag
\end{align}
The G-dynamics with respect to the ${{\hat{H}}^{\left( ri \right)}}$ is equal to
\begin{equation}\label{eq11}
  \hat{w}^{\left( ri \right)}=\frac{1}{\sqrt{-1}\hbar }{{\left[ s,{{\hat{H}}^{\left( ri \right)}} \right]}_{QPB}}={{b}_{c}}\left( 2u\frac{d}{dx}+u_{x}+2{{u}^{2}} \right)
\end{equation}where ${{b}_{c}}=-\frac{\sqrt{-1}\hbar }{2m}$. Conclusively, the one-dimensional G-dynamics of type I is then expressed as
$${{{\hat{w}}}^{\left( cl \right)}}=\frac{1}{\sqrt{-1}\hbar }{{\left[ s,{{\hat{H}}^{\left( cl \right)}} \right]}_{QPB}}={{b}_{c}}\left( 2u\frac{d}{dx}+u_{x}\right)$$
and type III is ${{w}^{\left( s \right)}}=2{{b}_{c}}{{u}^{2}}$, their summation produces
 the type II ${{{\hat{w}}}^{\left( ri \right)}}={{\hat{w}}^{\left( cl \right)}}+{{w}^{\left( s \right)}}$.
Definitely,  G-dynamics \eqref{eq11} in terms of $\hat{w}^{\left( ri \right)}$ is the equivalent expression of the \eqref{eq22} in one dimensional case.
Accordingly, the imaginary geomenergy follows
\[{{E}^{\left( \operatorname{Im} \right)}}\left( \hat{w}^{\left( ri \right)} \right)=\sqrt{-1}\hbar \hat{w}^{\left( ri \right)}=\frac{{{\hbar }^{2}}}{2m}\left( 2u\frac{d}{dx}+u_{x}+2{{u}^{2}} \right)\]Then it induces
 $$\sqrt{-1}\hbar {{\hat{w}}^{\left( ri \right)}}\psi=\frac{{{\hbar }^{2}}}{2m}\left( 2u{{\psi }_{x}}+\psi {{u}_{x}}+2\psi {{u}^{2}} \right)$$
Similarly, the covariant evolution of the geomentum operator is
\[{\mathcal{D}\over dt}{{\hat{p}}}^{{\left( ri\right)}}(t)={\sqrt{-1}\over \hbar }[\hat{H}^{\left( ri\right)},{{\hat{p}}}^{{\left( ri \right)}}(t)]\]
The calculation process is as follows
\begin{align}
 {{\left[ {{\hat{H}}^{\left( ri \right)}},{{\hat{p}}^{\left( ri \right)}} \right]}_{QPB}} & ={{\left[ V,{{\hat{p}}^{\left( ri \right)}} \right]}_{QPB}}=\frac{1}{2}m{{\omega }^{2}}{{\left[ {{x}^{2}},{{\hat{p}}^{\left( ri \right)}} \right]}_{QPB}}  \notag\\
 &=\sqrt{-1}\hbar m{{\omega }^{2}}x  \notag
\end{align}
The QGB in terms of ${{\hat{H}}^{\left( ri \right)}},{{\hat{p}}^{\left( ri \right)}} $ is
\[G\left( s,{{\hat{H}}^{\left( ri \right)}},{{\hat{p}}^{\left( ri \right)}} \right)={{\hat{H}}^{\left( ri \right)}}{{\left[ s,{{\hat{p}}^{\left( ri \right)}} \right]}_{QPB}}-{{\hat{p}}^{\left( ri \right)}}{{\left[ s,{{\hat{H}}^{\left( ri \right)}} \right]}_{QPB}}\]
where ${{\left[ s,{{\hat{p}}^{\left( ri \right)}} \right]}_{QPB}}=\sqrt{-1}\hbar u$ has been used, and the QGB in details becomes
\[G\left( s,{{\hat{H}}^{\left( ri \right)}},{{\hat{p}}^{\left( ri \right)}} \right)=\sqrt{-1}\hbar {{\hat{H}}^{\left( ri \right)}}u-{{\hat{p}}^{\left( ri \right)}}{{\left[ s,{{\hat{H}}^{\left( ri \right)}} \right]}_{QPB}}\]
As a consequence, the QCPB here is given by
\begin{align}
  \left[ {{\hat{H}}^{\left( ri \right)}},{{\hat{p}}^{\left( ri \right)}} \right]& ={{\left[ {{\hat{H}}^{\left( ri \right)}},{{\hat{p}}^{\left( ri \right)}} \right]}_{QPB}}+G\left( s,{{\hat{H}}^{\left( ri \right)}},{{\hat{p}}^{\left( ri \right)}} \right) \notag\\
 & =\sqrt{-1}\hbar m{{\omega }^{2}}x+\sqrt{-1}\hbar {{\hat{H}}^{\left( ri \right)}}u-{{\hat{p}}^{\left( ri \right)}}{{\left[ s,{{\hat{H}}^{\left( ri \right)}} \right]}_{QPB}} \notag
\end{align}
It turns out that the covariant evolution of the geomentum operator is derived as follows
\begin{align}
 \frac{\mathcal{D}}{dt}{{\hat{p}}^{\left( ri \right)}} &=\frac{\sqrt{-1}}{\hbar }\left[ {{\hat{H}}^{\left( ri \right)}},{{\hat{p}}^{\left( ri \right)}} \right]  =-m{{\omega }^{2}}x-{{\hat{H}}^{\left( ri \right)}}u+{{\hat{p}}^{\left( ri \right)}}\hat{w}^{\left( ri \right)}  \notag
\end{align} The G-dynamics seen from the \eqref{eq11} can evidently say how it forms and works associated with the geometric potential function.

As a result, the generalized Heisenberg equation with respect to the geomentum operator ${{\hat{p}}}^{{\left( ri \right)}}$ becomes
\begin{align}
 \frac{d}{dt}{{{\hat{p}}}^{\left( ri \right)}}\left( t \right) &=-m{{\omega }^{2}}x-\hat{H}^{\left( ri \right)}u \notag
\end{align}
To sum up, the differences between the quantum harmonic oscillator and generalized quantum harmonic oscillator are so clear, the mode is almost the same, this implies the QCPB can be a common calculation rules for quantum evolutions.

\section{Conclusions}
In this paper, we have proposed a series of new concepts based on the analysis of the G-dynamics included in the covariant dynamics defined by the QCHS in QCPB. On the solid foundation of the QCPB, we have gained some valuable consequences centred on the G-dynamics. By using the G-dynamics, we define new operators such as the D-operator, T-operator, G-operator, Ri-operator etc, we find their complex connections, in particular, we mainly focus on the their eigenvalues equation. With the help of the Schr\"{o}dinger equation, we analyze the G-dynamics and its related properties. We concretely discuss the different cases of  the G-dynamics of three types, in this way, we present the extension of the Schr\"{o}dinger equation, naturally. As an implication, we derive the generalized quantum harmonic oscillator and the covariant dynamics on it based on the QCHS theory.

\end{document}